\documentclass[
								reprint,
								aps,
								prl,
								nofootinbib,
								onecolumn,
								notitlepage,
							]{revtex4-1}

\usepackage{lmodern}
\usepackage{amsmath,amssymb,amsthm,amsfonts,mathrsfs}
\usepackage{graphicx,epstopdf,epsfig,enumerate}
\usepackage{color}
\usepackage[colorlinks=true,linkcolor=blue,citecolor=blue,urlcolor=blue]{hyperref}

\DeclareGraphicsExtensions{.eps}

\renewcommand{\vec}[1]{\mathbf{#1}}

\begin{document}

\title{Solar cell as a self-oscillating heat engine}

\author{Robert Alicki$^1$}
\email[e-mail: ]{fizra@ug.edu.pl}

\author{David Gelbwaser-Klimovsky$^2$}
\email[e-mail: ]{dgelbi@yahoo.com.mx}

\author{Krzysztof Szczygielski$^1$}
\email[e-mail: ]{fizksz@ug.edu.pl}
\affiliation{$^1$Institute of Theoretical Physics and Astrophysics, University of Gda\'nsk, 80-952 Gda\'nsk, Poland}
\affiliation{$^2$ Department of Chemistry and Chemical Biology, Harvard University, Cambridge, MA 02138, USA}

\begin{abstract}
Solar cells are  engines converting energy supplied by the  photon flux into work. All known types of macroscopic engines and turbines are also  self-oscillating systems which yield a periodic motion at the expense of a
usually non-periodic source of energy. The very definition of work in the formalism of quantum open systems suggests the hypothesis that the oscillating ``piston" is a necessary  ingredient of  the work extraction process. This aspect of solar cell operation is absent in the existing descriptions  and the main goal of this paper is to show that plasma oscillations provide the physical implementation of a piston. 
\end{abstract}

\maketitle

\section{Introduction}
The standard model of work extraction is based on an engine composed
of a working medium, a piston and two heat baths that are at equilibrium
with different temperatures. Its importance stems from its success
to set an universal bound to any work extraction process, the Carnot
bound, which shows, in agreement with the Kelvin's formulation of the
second law of thermodynamics, two different temperatures are needed
for work extraction. Besides, there is a complementary picture of an engine as a self-oscillating system \emph{``focusing on their ability to convert energy inputted at one
frequency (usually zero) into work outputted at another, well-defined frequency"} \cite{Jenkins}. This picture seems to apply to all types of turbines and motors \cite{Jenkins, AVK} and one can expect that it is equally correct for engines powered by a flux of photons. 
\par
The standard description of the photovoltaic cell involves the following processes \cite{solar_cells, Wuerfel}:\\ 
i) generation of the charge carriers due to the absorption of photons,\\
ii) subsequent separation of the photo-generated charge carriers in the junction.
\par
As noticed in \cite{Wuerfel} the often used explanation of the second process as caused by the emerging electric field in p-n junction cannot be correct. Charge separation is supposed to produce a DC current which, on the other hand, cannot be driven in a closed circuit by a purely electrical potential difference.
A standard thermodynamical explanation of electric current (work) generation in photovoltaic  and thermoelectric heat devices is the following :\\
Electrons gain energy in a form of heat current $J_H$ from the hot bath, then flow  against potential difference $\Phi$ producing useful power  $P= J_E \Phi$, where   $J_E$ is an electric current. A part of the heat described by the heat current $J_C$ is dumped to the cold bath.\\
The laws of thermodynamics put the following constraints
\begin{equation}
J_E \Phi = J_H - J_C
\label{ILaw}
\end{equation} 
\begin{equation}
\frac{J_H}{T_H} - \frac{J_C}{T_C} \leq 0 .
\label{IILaw}
\end{equation} 
Adding kinetic equations describing the processes of electron-hole creation, thermalization to the ambient temperature and recombination one obtains  the correct formulas for the open circuit voltage and  voltage-current relation. However, this picture does not explain the mechanism of persistent steady work extraction. Similarly, for a steam engine the net pressure due to the temperature difference obviously provides the net force acting on the piston but to explain the permanent periodic action of this engine we have to understand the details of operation of a piston linked to a flywheel and valves.
\par
This is exactly the place where the mechanism of self-oscillation supported  by the external constant energy flow enters into the game. 
In the following we propose a model in which plasma oscillations play a role of the periodic motion of a ``piston" and show that, indeed, under  realistic assumptions a positive power is supplied to this essentially classical macroscopic oscillator. Subsequently, the collective charge oscillations at THz frequencies are rectified by a p-n junction diode (``valve") to the output DC current.
\par
The mathematical formalism is based on the quantum Markovian master equations for slowly driven open quantum systems studied in \cite{Alicki:1979} (compare \cite{Gelbwaser:2013},\cite{Szczygielski:2013} for the fast driving case), and consistent with thermodynamics.
\section{Model of quantum engine}
We consider a model of  heat engine which consists of a ``working medium" called simply a system, two baths at different temperatures,  and a ``work reservoir" called often a ``piston" which is a system supplying to or extracting work from the working medium. Because work, in contrast to heat, is an ordered and deterministic form of energy we expect that a piston should be a macroscopic system operating in the semi-classical regime. Therefore, within the reasonable approximation can be replaced by external deterministic driving \cite{Gelbwaser:2013a}.
\par
For the readers convenience we briefly present the formalism of Quantum Master Equations (QME) based on the Davies  weak coupling limit \cite{Davies}, the Lindblad-Gorini-Kossakowski-Sudarshan generators and their extension to slowly varying external driving. Then the thermodynamical consequences are discussed and a special generic class of models with \emph{diagonal, weak driving}, applicable to the theory of solar cells, is presented.
\subsection{Master equation for open system with constant Hamiltonian}.

The total system consisting of a system  $\mathcal{S}$ weakly coupled to a bath $\mathcal{B}$. The total Hamiltonian is a sum of three terms:\\
i) a "bare" Hamiltonian of the system - $H^0$ replaced in the final formulas by  $H$ -  a``	physical renormalized Hamiltonian" containing the lowest order corrections,\\
ii) Hamiltonian of  the bath - $H_{\mathcal{B}}$,\\
iii) a system-bath (elementary) coupling:
\begin{equation}
H_{int} = A\otimes F\ ,\ \langle F\rangle_{\mathcal{B}} =0.
\label{}
\end{equation} 
where $A$ and $F$ are observables of the system and the bath, respectively. $\langle F\rangle_{\mathcal{B}}$ denotes the average with respect to the stationary state of the bath. 
\par
Two main ingredients enter the QME derived using Davies weak coupling limit procedure \cite{Davies}:\\
a) the spectral density of the bath
\begin{equation}
G(\omega)= \int_{-\infty}^{+\infty} e^{-i\omega t}\langle F(t)F\rangle_{\mathcal{B}}\, dt \geq 0,
\label{}
\end{equation} 
b) Fourier components of the coupling operator 
\begin{equation}
A(t) = U^{\dagger}(t) A U(t) = \sum_{\{{\omega}\}} A({\omega})e^{i {\omega}t}, \quad U(t) = e^{-iH t/\hbar}.
\label{}
\end{equation} 
Introducing system Hamiltonian spectral decomposition  and Bohr frequencies\\
\begin{equation}
{H} = \sum_k {\epsilon}_k |k\rangle\langle k, \quad \{{\omega}\}=\{({\epsilon}_k - {\epsilon}_l)/\hbar\}
\label{}
\end{equation} 
one obtains the relations 
\begin{equation}
[{H} ,A({\omega})] = -\hbar{\omega} A({\omega})\ ,\  A(-{\omega}) = {A^{\dagger}}({\omega}).
\label{transition}
\end{equation} 
where $A({\omega})$ called \emph{transition operators}  or \emph{Lindblad operators} correspond to energy exchange  of ${\hbar\omega}$.
The standard derivation yields the QME in the Schr\"odinger picture
\begin{equation}
{\frac{d\rho(t) }{dt}}=-\frac{i}{\hbar}[H , \rho(t)]+\mathcal{L}\rho(t)
\label{}
\end{equation} 
where
\begin{equation}
\mathcal{L}\rho = \frac{1}{2} \sum_{\{{\omega}\}}G({\omega})\bigl([A({\omega})\rho, A^{\dagger}({\omega})] + [A({\omega}),\rho A^{\dagger}({\omega})]\bigr)
\label{}
\end{equation} 
For  general  interactions $H_{int}= \sum_{\alpha} A_{\alpha}\otimes F_{\alpha}$
\begin{equation}
\mathcal{L}\rho = \frac{1}{2}\sum_{\alpha,\beta}\sum_{\{{\omega}\}}G_{\alpha\beta}({\omega})\bigl([A_{\alpha}({\omega})\rho, A_{\beta}^{\dagger}({\omega})] + [A_{\alpha}({\omega}),\rho A_{\beta}^{\dagger}({\omega})]\bigr)
\label{}
\end{equation} 
where $[G_{\alpha\beta}({\omega})]$ is a positively defined \emph{relaxation matrix}.

\subsection{Properties of QME }

The QME's obtained in the weak coupling limit possess the following properties \cite{Alicki:1976}, \cite{Alicki:1987}:

1) $\mathcal{L}$ possesses the \emph{Lindblad-Gorini-Kossakowski-Sudarshan}  structure what 
implies that the solution of QME is a \emph{completely positive, trace preserving, one-parameter semigroup}.

2) Hamiltonian part commutes with the dissipative one, i.e.
\begin{equation}
\rho(t) = \mathcal{U}(t)e^{t\mathcal{L}}\rho(0) = e^{t\mathcal{L}}\mathcal{U}(t)\rho(0), \quad\mathcal{U}(t)\rho \equiv U(t) \rho U^{\dagger}(t).
\label{}
\end{equation} 

3) Diagonal (in ${H}$ - basis) and off-diagonal density matrix elements evolve independently.

4) The stationary state $ \bar{\rho}$ satisfying $[H, \bar{\rho}]=0$ and $ \mathcal{L}\bar{\rho} = 0$, always exists for finite-dimensional Hilbert spaces.

5) For a single heat bath (i.e. a reservoir in the thermal equilibrium state):

a) spectral density satisfies the KMS condition
\begin{equation}
G(-\omega) = e^{-\hbar\omega/k_B T} G(\omega),
\label{}
\end{equation} 

b) the Gibbs state is stationary
\begin{equation}
{\rho}^{eq} = Z^{-1} \exp\Bigl\{- \frac{H}{k_B T}\Bigr\}, \quad \mathcal{L}{\rho}^{eq} = 0 ,
\label{}
\end{equation} 

c) the following quadratic form 
\begin{equation}
\langle X ,\mathcal{L}^*  Y \rangle_{eq} = \mathrm{Tr} \bigl( {\rho}^{eq} X^{\dagger} \mathcal{L}^* Y \bigr)
\label{Qform}
\end{equation}

is negatively defined \cite{Alicki:1976},\cite{Alicki:1987} , where $\mathcal{L}^* $ is the Heisenberg picture generator given by
\begin{equation}
\mathcal{L}^* X = \frac{1}{2}\sum_{\alpha,\beta}\sum_{\{{\omega}\}}G_{\alpha\beta}({\omega})\bigl(A_{\beta}^{\dagger}[X, A_{\alpha}({\omega})] +\bigl([A_{\beta}^{\dagger},X] A_{\alpha}({\omega})\bigr).
\label{Dbalance}
\end{equation} 
It means that the Heisenberg picture generator $\mathcal{L}^*$ can be treated as a hermitian, negatively defined operator acting on the space of (complex) observables equipped with the scalar 
product $\langle X ,  Y \rangle_{eq}\equiv  \mathrm{Tr} \bigl( {\rho}^{eq} X^{\dagger}  Y \bigr)$.
\subsection{Entropy balance and the Laws of Thermodynamics}
We identify the physical entropy with the von Neumann entropy of the reduced density matrix $S(\rho) = - k_B \mathrm{Tr}(\rho \ln\rho)$  and use also the \emph{ relative entropy} $ S(\rho | \sigma ) = k_B\mathrm{Tr}(\rho \ln\rho - \rho\ln\sigma)$.
\par
For the solution $\rho(t)$ of MME, and the stationary state  $\bar{\rho}$,  ($S(t)\equiv S(\rho(t))$)
\begin{equation}
\frac{d}{dt}S(t) =  \kappa(t) - k_B\frac{d}{dt} \mathrm{Tr}\bigl(\rho(t) \ln \bar{\rho}\bigr)
\label{}
\end{equation}
where $\kappa(t)  = - \frac{d}{dt} S(\rho(t)|\bar{\rho})= - k_B\mathrm{Tr}\bigl([\mathcal{L}\rho(t)][\ln\rho(t) - \ln \bar{\rho}]\bigr)\geq 0$ is interpreted as an \emph{entropy production}. Positivity of the entropy production follows from the fact that for any completely positive and trace-preserving map $\Lambda$,  $S(\Lambda\rho |\Lambda \sigma ) \leq S(\rho | \sigma )$ \cite{Spohn:1978}.
For many independent heat baths one obtains the Second Law in the following form

\begin{equation}
\frac{d S}{dt} -  \sum_{k} \frac{J_k}{T_k} \geq 0 , 
\label{}
\end{equation} 
where $J_k$ is a heat current flowing from the $k$-th bath.
\subsection{The case of a slow piston}
In order to define work we introduce the time-dependent  Hamiltonian $H(t) = H_0 + V(t)$ with slowly varying and typically periodic perturbation $V(t)$ which gives a semi-classical description of a piston.  We combine now the weak coupling assumption concerning the interaction of the system with several baths with a kind of adiabatic approximation concerning the time-dependent driving. The former condition means, practically, that the relaxation rates are much smaller than the corresponding Bohr frequencies. The later is valid for the case when the time scale of driving is much slower than the time scale determined by the relevant Bohr frequencies. This is a similar situation to standard adiabatic theorem in quantum mechanics and implies that in the derivation of Master equation we can put the temporal values of Bohr frequencies $\{\omega(t)\}$ and transition operators  $\{A_{\alpha}({\omega(t)})\}$ satisfying \eqref{transition} with $H$ replaced by $H(t)$.
\par	
Under the conditions of above one obtains the following form of QME
\begin{equation}
\frac{d}{dt}\rho(t) = -\frac{i}{\hbar}[H(t),\rho(t)] + \sum_j \mathcal{L}_j(t)\rho(t),
\label{}
\end{equation}
where $\mathcal{L}_j(t)$ is a LGKS generator obtained by a weak coupling to the $j$-th bath and for a fixed $H(t)$.
The properties of generators $\mathcal{L}_j(t)$ imply the \emph{Zero-th Law of Thermodynamics} ($\beta_j = 1/k_B T_j$)
\begin{equation}
\mathcal{L}_j(t)\rho_j^{eq}(t)=0\  ,\ \rho_j^{eq}(t) = \frac{e^{-\beta_j H(t)}}{\mathrm{Tr}e^{-\beta_j H(t)}}\ .
\label{eq}
\end{equation} 
Using the definitions \cite{Alicki:1979}:
$W$-work provided by $\mathcal{S}$, $Q$ - heat absorbed by $\mathcal{S}$, $E$ - internal energy of $\mathcal{S}$
\begin{equation}
E(t) = \mathrm{Tr}\bigl(\rho(t)H(t)\bigr) 
\label{intenergy}
\end{equation}
\begin{equation} 
\frac{d}{dt}W(t) = -\mathrm{Tr}\bigl(\rho(t)\frac{dH(t)}{dt}\bigr)\ ,
\label{Swork}
\end{equation} 
\begin{align}
 \frac{d}{dt}Q(t) &= \mathrm{Tr}\bigl(\frac{d\rho(t)}{dt}H(t)\bigr)\\
&= \sum_j\mathrm{Tr}\bigl(H(t)\mathcal{L}_j(t)\rho(t)\bigr)\equiv \sum_j\frac{d}{dt}Q_j(t) ,
\label{heat}
\end{align}
where $Q_j$ is a heat absorbed by $\mathcal{S}$ from  $\mathcal{B}_j$,
one obtains the \emph{ First Law of Thermodynamics}
\begin{equation} 
\frac{d}{dt}E(t) = \frac{d}{dt}Q(t) - \frac{d}{dt}W(t) .
\label{}
\end{equation} 
The \emph{ Second Law of Thermodynamics} follows again from Spohn innequality 
\begin{equation}
\frac{d}{dt}S(t) - \sum_j\frac{1}{T_j} \frac{d}{dt}Q_j(t)= \sum_j\sigma_j(t)\geq 0 
\label{SIIlaw}
\end{equation}
where $\sigma_j(t)$ is an entropy production caused by $\mathcal{B}_j$ and given by
\begin{equation}
\sigma_j(t)=  k_B \mathrm{Tr}\bigl(\mathcal{L}_j(t)\rho(t)[\ln \rho(t) - \ln \rho_j^{eq}(t)]\bigr)\geq 0.
\label{Senprod}
\end{equation} 
\subsection{Weak, diagonal and periodic driving}
We consider a generic case of oscillating weak driving $V(t)$ which in the lowest order approximation can be replaced by the diagonal (in the basis of $H_0$) operator
\begin{equation}
V(t) = g M \sin\Omega t ,\quad [H_0 , M]=0,
\label{M-perturbation}
\end{equation} 
with the small coupling constant $g << 1$. In this case all Hamiltonians $H(t)$ commute.
\par
The unitary part of the dynamics $U(t)$ governed by $H(t)$ commutes with $H_0$ and $M$. One can write  $\mathcal{L}(t) =\mathcal{L}[\xi(t)] $ 
where $\mathcal{L}[\xi]$ is  computed with the system Hamiltonian $H_0 + \xi M$ and $\xi(t) = g \sin\Omega t$. Again for different $\xi$'s the super-operators  $\mathcal{L}[\xi]$  commute with the Hamiltonian part $-i[H(t) , \cdot]$ and one can use their lowest order expansion with respect to $\xi$
\begin{equation}
 \mathcal{L}[\xi] = \mathcal{L}[0] + \xi \mathcal{L}'[0] + \mathcal{O}(\xi^2),
\label{Slowest_order}
\end{equation}
In the next step we use the lowest order expression for the dissipative part of the super-propagator
\begin{align}\label{Slowest_order1}
\Lambda_D(t) &= \exp\bigl\{\int_0^t \mathcal{L}[\xi(s)] ds\bigr\}  \\
&\simeq e^{t\mathcal{L}[0]} + g\int_0^t (\sin \Omega s)\, e^{(t-s)\mathcal{L}[0]} \mathcal{L}'[0]e^{s\mathcal{L}[0]}\, ds. \nonumber
\end{align}
Applying  now the definition of work \eqref{Swork}, using the commutation properties of the generator $\mathcal{L}[\xi]$, and  the fact that $[U(t) ,M ]= 0$,  one obtains the formula for the stationary average power output 
\begin{align}\label{Spower}
\bar{P} &\equiv -\lim_{t_0\to\infty}\frac{1}{t_0}\int_0^{t_0}\mathrm{Tr}\bigl(\rho(t)\frac{dH(t)}{dt}\bigr)\, dt \\ 
&= -g\Omega\lim_{t_0\to\infty} \frac{1}{t_0}\int_0^{t_0}\mathrm{Tr}\Bigl(M\Lambda_D(t)\rho(0) \Bigr)\cos\Omega t\,dt\nonumber .
\end{align}
Inserting the lowest order expression \eqref{Slowest_order1} into \eqref{Spower} and using the fact that all super-operators commute one can compute the super-operator-valued integral like a usual one. Then, we apply the obvious assumption that the unperturbed dynamics $e^{t\mathcal{L}[0]} $ drives, asymptotically, any initial state $\rho(0)$ to the stationary state denoted by $\bar{\rho}[0]$. Finally, the limit $t_0\to\infty$ in \eqref{Spower} can be performed leading to the second order  approximation for the average output power
\begin{equation}
\bar{P} = \frac{1}{2}g^2 \mathrm{Tr}\Bigl(M \frac{\Omega^2}{\Omega^2 + (\mathcal{L}[0])^2}\mathcal{L}'[0]\bar{\rho}[0]\Bigr) .
\label{Spower1}
\end{equation}
$\mathcal{L}[\xi]$ also possesses the stationary state $\bar{\rho}[\xi]$, i.e. $\mathcal{L}[\xi] \bar{\rho}[\xi] = 0$ and hence we can use the identity 
\begin{equation}
\mathcal{L}'[\xi]\bar{\rho}[\xi] = - \mathcal{L}[\xi]\bar{\rho}'[\xi]
\label{identity}
\end{equation}
where ``prime" denotes the derivative with respect to $\xi$. Then, replacing the Schroedinger picture generator $\mathcal{L}[0]$ by the Heisenberg picture one $\mathcal{L}^*[0]$ we can transform the formula \eqref{Spower1} into
\begin{equation}
\bar{P} = -\frac{1}{2}g^2 \mathrm{Tr}\Bigl(\bar{\rho}'[0] \frac{\Omega^2}{\Omega^2 + ({\mathcal{L}^*}[0])^2}\mathcal{L}^*[0] M\Bigr) .
\label{Spower2}
\end{equation}
For the case when the decay rate of $M$ is much lower than the modulation frequency we can neglect $({\mathcal{L}^*}[0])^2$ in \eqref{Spower2} to obtain the simplified expression 
\begin{equation}
\bar{P} = -\frac{1}{2}g^2 \mathrm{Tr}\Bigl(\bar{\rho}'[0] \mathcal{L}^*[0] M\Bigr) .
\label{Spower3}
\end{equation}
The compact formula of above will be used to derive the specific expression for the solar cell power.
\subsection{No output power from a single heat bath}
The obtained lowest order formulas for power \eqref{Spower2}, \eqref{Spower3} are still consistent with thermodynamics. Namely, assuming that the reservoir is a thermal equilibrium bath  at the  temperature $T$ we have the following properties:
\begin{enumerate}[1)]
\item $\bar{\rho}[\xi]$ is the Gibbs state with respect to the Hamiltonian $H_0 +  \xi M$,
\item the Heisenberg picture generator $\mathcal{L}^*[0]$ is a negatively defined operator on the Hilbert space equipped with the scalar product $\langle X ,  Y \rangle_{eq}\equiv  \mathrm{Tr} \bigl( \bar{\rho}[0] X^{\dagger}  Y \bigr)$ (compare with \eqref{Qform}, \eqref{Dbalance}).
\end{enumerate}
Using the first property one can compute $\bar{\rho}'[0]$ and rewrite the formula \eqref{Spower2} as
\begin{align}\label{Seq}
\bar{P}_{eq} &= \frac{g^2}{2 k_B T} \mathrm{Tr}\Bigl(\bar{\rho}[0]M \frac{\Omega^2}{\Omega^2 + ({\mathcal{L}^*}[0])^2}\mathcal{L}^*[0] M\Bigr)\\ 
&= \frac{g^2}{2 k_B T} \langle M ,\frac{\Omega^2}{\Omega^2 + ({\mathcal{L}^*}[0])^2}\mathcal{L}^*[0] M\rangle_{eq} \leq 0
\label{Spower4}
\end{align}
which is obviously negative as well as its simplified version \eqref{Spower3}. Therefore, as expected, one cannot extract power from a single heat bath.
\subsection{Feed-back mechanism}
For a reservoir composed of two equilibrium ones at different temperatures a positive output power $\bar{P} >0$ can be obtained and
in this case the mechanism of self-oscillations works. Treating the external perturbation as caused by a coupling of the system to a macroscopic oscillator we see that the positive power is supplied to the oscillator increasing the amplitude of its oscillation (positive feed-back) up to the moment when the net energy supply from the reservoir is compensated by the load attached to the oscillator.

\section{Model of semiconductor solar cell}
\begin{figure}[htbp]
	\centering
		\includegraphics{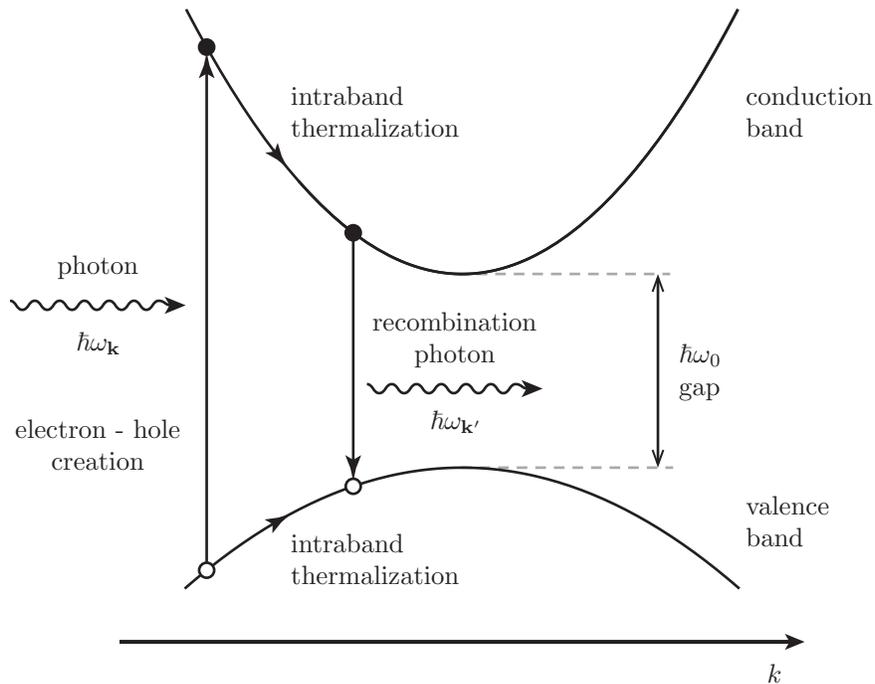}
	\caption{Schematic picture of leading processes involving electrons, photons and phonons in a semiconductor with a direct band.}
	\label{fig:Figure4}
\end{figure}

A solar cell is an engine which produces work from heat exchanged  with  a non-equilibrium bath. The bath consists of the photonic non-equilibrium reservoir  characterized by the local state population $n[\omega]$ and the basically phononic heat bath at the temperature $T$ of the device.
The typical semiconductor solar cell consists of a moderately doped p-type  absorber,  on both sides of which a highly doped layer is formed, n-type on
the top side and p-type on the back side, respectively.  The electronic states in the valence band and in the conduction band are labeled by the index $\mathbf{k}$ which corresponds to the quasi-momentum $\hbar\mathbf{k}$ (spin can be easily added) with the energies $E_v(\mathbf{k})$ and $E_c(\mathbf{k})$ . We assume a direct band structure with vertical optical transitions which preserve quasi-momentum (see fig. \ref{fig:Figure4}). 
\par
The basic irreversible processes are the following:
\begin{enumerate}[i)]  
\item fast intraband thermalization processes mediated by phonons and described by  $\mathcal{L}_{th}$.
\item optical transitions between the valence and conduction band, which create or annihilate electron-hole pairs, described by  $\mathcal{L}_{em}$,
\item non-radiative electron-hole recombination  which is neglected in our idealized model.\\
\end{enumerate}
\subsection{Plasma oscillations and current rectification }
The fundamental question in the presented approach to work  generation in solar cell is the origin of periodic oscillations which can be seen as classical. The frequency $\Omega$ is assumed in our derivations to be much smaller than the frequency $\omega_0$, but much larger than the recombination rate in order to justify \eqref{Spower3}. The only phenomenon which satisfies all these requirements is plasma oscillation visible for p-n junctions. Their appearance is due to the fact  that a p-n junction creates an interface between regions of different electron concentrations which can oscillate in space producing collective macroscopic electric field oscillations.  In several experiments such oscillations have been observed  \cite{plasma1,plasma2}, with  typical frequencies $\Omega/2\pi \simeq 1 THz$, much lower than $\omega_0 \simeq 1400 THz$ corresponding to the energy gap $\sim 1 eV$. On the other hand $\Omega$ is much higher than the recombination rate $\sim 10^{4} s^{-1}$ what justifies the transition from \eqref{Spower2}  to \eqref{Spower3}.
\begin{figure}[htbp]
	\centering
		\includegraphics{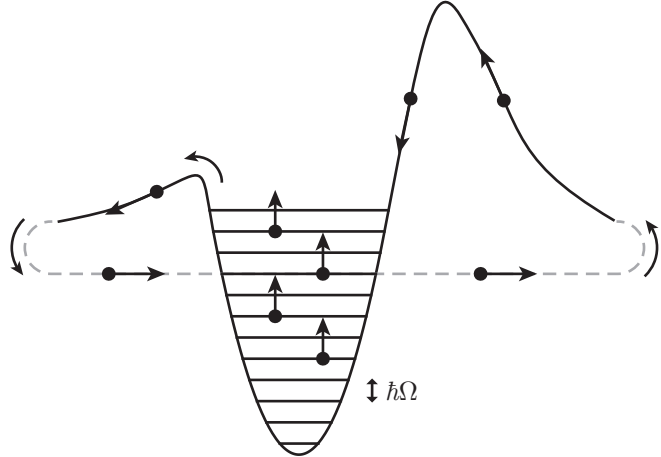}
	\caption{{\bf Schematic picture of rectification of plasma oscillations.} Plasma oscillations caused by the collective motion of free carriers represented by the dimensionless amplitude $\xi(t)$ in the eq.\eqref{ham_pert1}. For small $\xi$ plasma self-oscillation is described by the quantum harmonic oscillator coherently pumped by the feed-back mechanism.  Due to the asymmetric diode-type potential at the edges, charge oscillations are converted into a DC current.} 
	\label{fig:Figure5}
\end{figure}
In the final step of cell operation the THz plasma oscillations must be converted into a direct current.  A qualitative picture of this mechanism is shown on Fig. \ref{fig:Figure5}. The plasmonic degree of freedom is represented by the quantum levels in the asymmetric potential which is harmonic for lower energies. Asymmetry is due to the p-n junction which defines an ``easy" direction for the carrier flow (to the left). The work supplied to the oscillator drives the unidirectional electric current.\\
\subsection{Hamiltonians, Master equations and stationary states}
The electrons in a semiconductor occupying the conduction and valence bands are described by the annihilation and  creation operators $c_{\vec{k}}$, $c^{\dagger}_{\vec{k}}$ and $v_{\vec{k}}$, $v^{\dagger}_{\vec{k}}$, respectively, subject to canonical anticommutation relations. The unperturbed Hamiltonian   reads
\begin{equation}
H_{0} =  \sum_{\vec{k}} \Bigl(E_{c}(\vec{k}) c^{\dagger}_{\vec{k}} c_{\vec{k}} + E_{v}(\vec{k}) v^{\dagger}_{\vec{k}} v_{\vec{k}}\Bigr). 
\label{ham_electrons}
\end{equation}
In the p-n junction a non-homogeneous free carrier distribution created in a self-consistent build-in potential  can be perturbed producing collective plasma oscillations with the frequency $\Omega$. The associated time-dependent perturbation  added to the electronic Hamiltonian \eqref{ham_electrons} has  a mean-field form  ($N$ - number of atoms in the sample)
\begin{equation}
\xi(t) M = \xi(t)\frac{1}{\sqrt{N}} E_g \sum_{\vec{k}} \bigl( c^{\dagger}_{\vec{k}} c_{\vec{k}} +  v_{\vec{k}}v^{\dagger}_{\vec{k}} \bigr), 
\label{ham_pert1}
\end{equation}
where $\xi$ is a small dimensionless parameter describing the magnitude of deformation, $E_g$ is the relevant energy scale, and  $c^{\dagger}_{\vec{k}} c_{\vec{k}}$, $ v_{\vec{k}}v^{\dagger}_{\vec{k}}$ are number operators of free electrons and holes, respectively.
\par
To apply the formulas derived in the previous sections we notice first that the driving perturbation \eqref{ham_pert1} depends only on the total numbers of both types of carriers and hence does not interfere with intraband transitions. Therefore, the relevant Bohr frequency is associated with the gap $E_g$ yielding the time scale $\sim 10^{-15} s$, much faster than the modulation period $\sim 10^{-12} s.$ This justifies the adiabatic approximation.  The weak coupling assumption is for sure satisfied for very slow radiation recombination processes.
\par
Among the basic irreversible processes the intraband thermal relaxation is the fastest (thermalization time $\sim 10^{-12} s$) and therefore, the stationary state of the electronic systems with the total Hamiltonian $H_0 + \xi M$ is, within a reasonable approximation, a product of grand canonical ensembles for electrons in conduction and valence band with the same temperature $T$ of the device
and different electro-chemical potentials $\mu_c$ and $\mu_v$, respectively. The associated density matrix has form
\begin{align}
\bar{\rho}[\xi] = &\frac{1}{Z[\xi]} \exp\left\{ -\frac{1}{k_B T} \sum_{\vec{k}} \left[ \left(E_c(\vec{k})+ \frac{\xi E_g}{\sqrt{N}} -\mu_c\right) c^{\dagger}_{\vec{k}}c_{\vec{k}} \right. \right. \nonumber \\
&\left. \left. - \left(E_v(\vec{k}) - \frac{\xi E_g}{\sqrt{N}}-\mu_v\right) v_{\vec{k}}v^{\dagger}_{\vec{k}}\right]\right\} . 
\label{grand}
\end{align}
The electro-chemical potentials are determined by the numbers of carriers and hence by doping and  radiative and non-radiative processes of electron-hole creation and recombination. 
\par
Because the intraband thermalization to the ambient temperature $T$ does not change the number of free electrons and holes, i.e. $\mathcal{L}^*_{th}M = 0$, the generator $\mathcal{L}^*_{th}$ does not enter the formula for power \eqref{Spower3}. Here, one can doubt whether for such fast relaxation the weak coupling condition and hence the validity of the Markovian approximation leading to $\mathcal{L}_{th}$ holds. However, intraband relaxation does not contribute to work generation but only determines the structure of the stationary state. The form of this state expressed in terms of Fermi-Dirac distributions is generally accepted in the literature \cite{Wuerfel} and the accuracy of the Markovian approximation for the thermalization process is not very relevant.  Finally, the contribution which remains in the eq. \eqref{Spower3} describes the quasi-momentum preserving (vertical) transitions and reads
\begin{align}
		\mathcal{L}_{em}[0] \rho &= \frac{1}{2} \sum_{\vec{k}}\Bigl\{\gamma_{rec}(\vec{k})\bigl( [c_{\vec{k}} v_{\vec{k}}^{\dagger} , \rho \,v_{\vec{k}} c_{\vec{k}}^{\dagger}] + [c_{\vec{k}} v_{\vec{k}}^{\dagger}\, \rho ,v_{\vec{k}} c_{\vec{k}}^{\dagger}] \bigr) \nonumber \\
		&+\gamma_{ex}(\vec{k}) \bigl([ c_{\vec{k}}^{\dagger} v_{\vec{k}}, \rho \,  v_{\vec{k}}^{\dagger} c_{\vec{k}}] +[ c_{\vec{k}}^{\dagger} v_{\vec{k}}\, \rho ,  v_{\vec{k}}^{\dagger} c_{\vec{k}}] \bigr)\Bigr\}, 
		\label{ME_solar}
	\end{align}
\begin{equation}
\gamma_{rec}(\vec{k}) =  \frac{1}{\tau_{se}} \bigl[1 + n (\omega_{\vec{k}})\bigr] , 
 \gamma_{ex}(\vec{k}) =  \frac{1}{\tau_{se}}  n (\omega_{\vec{k}})
\end{equation}
where   $\tau_{se}$ is the spontaneous emission time, $\hbar\omega_{\vec{k}} = E_c(\vec{k})- E_v(\vec{k}) $, and $n(\omega)$ denotes a number of photons occupying a state with the frequency $\omega$.
\subsection{Power and efficiency}
One can insert all elements computed in the previous section into the expression for power \eqref{Spower3}. Then we use the properties of the quasi-free (fermionic Gaussian) stationary state \eqref{grand} which allow to reduce the averages of even products of annihilation and creation fermionic operators into sums of products of the Fermi-Dirac distribution functions 
\begin{equation}
f_{c}(\vec{k}) = \frac{1}{e^{\beta (E_{c}(\vec{k})-\mu_{c})}+1}, \quad
f_{v}(\vec{k}) = \frac{1}{e^{\beta (E_{v}(\vec{k})-\mu_{v})}+1},
\end{equation}
with $\beta = 1/k_B T$. 
\par
The leading order contribution to power possesses  a following form
\begin{align}\label{power3a}
\bar{P} = \frac{g^2  E_g^2}{k_B T}\frac{N_{car}}{N}\sum_{\vec{k}}\Bigl( &\gamma_{ex}(\vec{k}) \bigl[ 1 -f_{c}(\vec{k})\bigr] f_{v}(\vec{k}) \\
&- \gamma_{rec}(\vec{k}) \bigl[ 1 -f_{v}(\vec{k})\bigr] f_{c}(\vec{k})\Bigr), \nonumber
\label{power3a}
\end{align}
where $N_{car}= \sum_{\vec{k}}\bigl[\langle c^{\dagger}_{\vec{k}} c_{\vec{k}} \rangle_0 + \langle v_{\vec{k}}v^{\dagger}_{\vec{k}}\rangle_0 \bigr]$
is the total number of free charge carriers.
\par
One can introduce the local temperature of light $T[\omega]$ defined by 
\begin{equation}
e^{-{\hbar\omega}/k_B T[\omega]} = \frac{n (\omega)}{1 + n (\omega)} . 
\label{local_temp}
\end{equation}
For the incident sunlight on Earth a rough approximation holds
\begin{equation}
n_{sun}(\omega) = \frac{\lambda}{e^{\hbar\omega/k_B T_s}-1} ,
\label{rough}
\end{equation}
where $T_s\simeq 6000 K$ is the temperature of the Sun surface and $\lambda = [R_{sun}/R_0]^2\simeq 2\times 10^{-5} $ is the geometrical factor ($R_{sun}$ - Sun radius) which takes into account the photon density reduction at large distance from the source. In particular, for the typical value of the energy gap $E_g = \hbar\omega_0 \simeq 1 eV$ the effective temperature of sunlight $T[\omega_0] \simeq 1000 K $. 

Because  the product of population numbers for free carriers given by $\bigl[ 1 -f_{v}(\vec{k})\bigr] f_{c}(\vec{k})$ is essentially concentrated on the interval $\omega_{\vec{k}} \in [\omega_0 , \omega_0 + \mathcal{O}(k_B T/\hbar)]$, and  $ T\simeq 300K \ll T[\omega_0]\simeq 1000 K \ll \hbar\omega_0/k_B \simeq 12 000 K$, the expression \eqref{power3a} can be approximated by
\begin{equation}\label{power4}
\bar{P} = \frac{g^2  E_g^2 N_{car}\bar{F}}{k_B T\, \tau_{se}} \times 
\Bigl(\exp\Bigl\{\frac{1}{k_B T}\Bigl(\Bigl[ 1-\frac{T}{T[\omega_0]}\Bigr]\hbar\omega_0 - e\Phi\Bigl)\Bigr\} -1\Bigr)
\end{equation}
where $\bar{F} =\frac{1}{N}\sum_{\vec{k}}\bigl[1 + n (\omega_{\vec{k}})\bigr]\bigl[ 1 -f_{v}(\vec{k})\bigr] f_{c}(\vec{k}) > 0$ is independent of the size of a cell, and the \emph{voltage} $\Phi$ is identified with the difference of electro-chemical potentials, i.e. $e\Phi \equiv \mu_{c} - \mu_{v}$.\\
The condition for work generation by the solar cell reads
\begin{equation}
e\Phi <  e\Phi_0 = \eta_C\,E_g ,\quad \eta_C =  1-\frac{T}{T[\omega_0]}
\label{positive_work}
\end{equation}
what implies that $\Phi_0$ is an \emph{open-circuit voltage} of the cell for the idealized case \cite{Carnot}. 
\par
The presence of the Carnot factor $\eta_C$ suggests also the interpretation of the eq. \eqref{positive_work} in terms of thermodynamical efficiency. Indeed, the incident photon of the frequency $\omega > \omega_0$ produces an excitation of the energy close to $E_g$ in the process of electron-hole creation followed by the fast thermalization of an electron to the bottom of the conduction band, and a hole to the top of the valence one. Then, a part of energy $E_g $ is transformed into useful work, equal at most $e\Phi_0$ per single electron flowing in the external circuit.   
The maximal efficiency $\eta_{max}$ under the conditions that each photon with the energy higher than the gap produces an electron-hole pair and  non-radiative recombination processes are neglected, is given by the product
$
\eta_{max} = \eta_{u}\cdot \eta_C,
$
where $\eta_u$ is the so-called \emph{ultimate efficiency} computed under the assumptions:\\
 a) \emph{``... photons with energy
greater than $E_g$ produce precisely the same effect as photons of energy $E_g$, while photons of lower energy
will produce no effect"} \cite{Shockley},\\
 b) the whole $E_g$ is transformed into work. 
\par
Under standard illumination conditions the ultimate efficiency of a solar cell can reach $44\%$ and the Carnot factor is about $70\%$ what yields $\eta_{max}\simeq 31\% $ - the Shockley's \emph{detailed balance limit} \cite{Shockley}.
Actually, photons are absorbed along their path in the absorber and $n(\omega_0)$ decays exponentially with the penetration distance. Taking a more realistic  average value $\bar{T} = (T[\omega_0] + T)/2 \simeq 650 K$ one obtains $\eta_{max}\simeq 24\% $, much closer to the real performance of standard $GaAs$ solar cells.
\subsection{Conclusions} 
The presented model based on the idea of self-oscillations explains the dynamical origin of work generation in photovoltaic cells which is not present in the standard ``static" picture. The main new ingredient is the role of plasma oscillation as a ``piston" which transforms the steady heat input from the photon flux into periodic motion. This model provides a bridge between the theory of driven quantum  open systems applied to heat engines and the theory of photovoltaic devices. The formulas \eqref{local_temp} and \eqref{positive_work} explain in a simple way the meaning of the ``light temperature", the Carnot bound,  and the linear relation between the open circuit voltage and the device temperature. 
\par
The experimental verification of this model should provide the evidence of THz plasma oscillation in the device with the amplitude square proportional to the power output. Such oscillations produce a weak THz radiation which, in principle, could be detected.
\par
The similar ideas can be applied to  other types of heat engines with ``hidden self-oscillations". It seems that thermoelectric devices based either on bimetallic or semiconductor p-n junctions can be described by the very similar models. Plasma oscillation remains a piston and sunlight is replaced by the hot bath. For organic photovoltaic systems, proton pumps or photosynthesis there exists quite strong evidence of the important role of coherent molecular oscillations played in the energy and charge transfer (see e.g. \cite{Yakovlev}). It is plausible that those oscillations can play the role of a piston in the work extraction mechanism as well.

\subsection{Acknowledgments} R.A. and K.S. are supported by the Foundation for Polish Science TEAM project co-financed by the EU European Regional Development Fund. D.G-K is supported by CONACYT.

\end{document}